\documentclass[aps,twocolumn,showpacs]{revtex4}
\usepackage{epsfig}
\newcommand{\be}{\begin{equation}}
\newcommand{\ee}{\end{equation}}
\newcommand{\ba}{\begin{array}}
\newcommand{\ea}{\end{array}}

\begin{document}
\title{Metropolis simulations of Met-Enkephalin with\\ 
solvent-accessible area parameterizations}
\author{Bernd A. Berg}
\affiliation{Department of Physics, Florida State University,
Tallahassee, FL 32306}
\affiliation{School of Computational Science and
Information Technology, Florida State University,
Tallahassee, FL32306}
\author{Hsiao-Ping Hsu}
\affiliation{John-von-Neumann Institute for Computing, Forschungszentrum
J\"ulich, D-52425 J\"ulich, Germany}

\date{\today}

\begin{abstract}
We investigate the solvent-accessible area method by means of Metropolis 
simulations of the brain peptide Met-Enkephalin at 300$\,K$. For the 
energy function ECEPP/2 nine atomic solvation parameter (ASP) sets are 
studied. The simulations are compared with one another, with simulations
with a distance dependent electrostatic permittivity $\epsilon (r)$, and 
with vacuum simulations ($\epsilon =2$). Parallel tempering and the 
biased Metropolis techniques RM$_1$ are employed and their performance
is evaluated. The measured observables include energy and dihedral 
probability densities (pds), integrated autocorrelation times, and 
acceptance rates. Two of the ASP sets turn out to be unsuitable for 
these simulations. For all other systems selected configurations are 
minimized in search of the global energy minima, which are found for 
the vacuum and the $\epsilon(r)$ system, but for none of the ASP models.
Other observables show a remarkable dependence on the ASPs. In particular, 
we find three ASP sets for which the autocorrelations at 300$\,$K are 
considerably smaller than for vacuum simulations.
\end{abstract}


\maketitle

\section{Introduction}

In nature biomolecules exist in the environment of solvents, thus the 
molecule-solvent interactions must be taken into account. It is very 
computer time consuming to simulate models for which the molecules of 
the surrounding water are treated explicitly. Therefore, a number of 
approximations of solvent effects have been developed. In the 
solvent-accessible area approach~\cite{lr71,c74,em86} it is assumed 
that the protein-solvent interaction is given by the sum of the surface 
area of each atomic group times the atomic solvation parameter (ASP).
The choice of a set of ASPs (also called hydrophobicity parameters or
simply hydrophobicities) defines a model of solvation.  However, there 
is no agreement on how to determine the universally best set of ASPs, 
or a at least the best set for some limited purpose. For instance, eight 
sets were reviewed and studied by Juffer et al.~\cite{jeh95} and it was 
found that they give rather distinct contributions to the free energy 
of proteins folding. 

In this paper we investigate how different ASP sets modify the
Metropolis simulations of the small brain peptide Met-Enkephalin 
(Tyr-Gly-Gly-Phe-Met) at 300$\,$K. The reason for the choice of 
Met-Enkephalin is that its vacuum properties define a reference 
system for testing numerical methods, 
e.g.~\cite{ls87,okk92,ho93,mmm94,hoo99,b03}. 
Therefore, Met-Enkephalin appears to be well suited to set 
references for the inclusion of solvent effects as well, but we 
are only aware of few articles~\cite{ls88,koh98}, 
which comment on the modifications due to including a solvent model. 

We set our simulation temperature to 300$\,$K, because room temperature 
is the physical temperature at which biological activity takes 
place. Most of the previous simulations of Met-Enkephalin in vacuum 
were performed at much lower temperatures or employed elaborate 
minimization techniques with the aim to determine the global energy 
minimum (GEM). Only recently~\cite{b03} it was shown that the GEM is 
well accessible by local minimization of properly selected configurations 
from an equilibrium time series at 300$\,$K. Precisely this should be the 
case for a GEM which is of relevance at physical temperatures. 

For our simulations we use the program package SMMP~\cite{smmp} 
(Simple Molecular Mechanics for Proteins) together with parallel
tempering~\cite{g91,hn96,ha97} (PT) and the recently introduced~\cite{b03} 
biased Metropolis technique RM$_1$ (rugged Metropolis -- 
approximation~1). SMMP implements a number of all-atom energy 
functions, describing the intramolecular interactions, and nine 
ASP sets~\cite{em86,oons,sch1,sch2,jrf,we92,sch4,bm} to model the
molecule solvent interactions. We use the ECEPP/2~\cite{sns84} 
(Empirical Conformational Energy Program for Peptides) energy 
function with fully variable $\omega$ angles and
simulate all nine ASP sets. For comparison we simulate also 
Met-Enkephalin in vacuum and with the distance dependent 
electrostatic permittivity $\epsilon (r)$ of Ref.~\cite{hrf85}.

  The paper is organized as follows: The energy functions and Metropolis
methods used are explained in Sec.~\ref{basics}. In Sec.~\ref{results} we 
present our results from simulations of the brain peptide Met-Enkephalin. 
Summary and conclusions are given in Sec.~\ref{conclusions}.

\section{Models and Methods} \label{basics}

\subsection{ASP sets}

In all-atom models of biomolecules the total conformational energy of 
the intramolecular interactions $E_I$ is given as the sum of the 
electrostatic, the Lennard-Jones (Van der Waals), the hydrogen bond, 
and the torsional contributions,
\begin{eqnarray}
E_I\ =\ 332\sum_{i<j} \frac{q_i q_j}{\epsilon r_{ij}}\ +\ \sum_{i<j} 
  \left(\frac{A^{\rm LJ}_{ij}}{r_{ij}^{12}}  \nonumber  
- \frac{B^{\rm LJ}_{ij}}{r_{ij}^6}\right)\ +\\ 
  \sum_{i<j} \left(\frac{A^{\rm HB}_{ij}}{r_{ij}^{12}}
- \frac{B^{\rm HB}_{ij}}{r_{ij}^{10}}\right)\ +\ 
\sum_k U_k\, [1\pm \cos(n_k \phi_k)] \, . \label{eceep2}
\end{eqnarray}
Here $r_{ij}$ is the distance between atoms $i$ and $j$, $q_i$ and $q_j$ 
are the partial charges on the atoms $i$ and $j$, $\epsilon$ is the 
electric permittivity of the environment, $A_{ij}$, $B_{ij}$, $C_{ij}$ 
and $D_{ij}$ are parameters that define the well depth and width for a 
given Lennard-Jones or hydrogen bond interaction, and $\phi_k$ is the 
$k$th torsion angle. The units are as follows: distances are in \AA,
charges are in units of the electronic charge and energies are in
kcal/mol.

One of the simplest ways to include interactions with water is to
assume a distance dependent electrostatic permittivity according to
the formula~\cite{hrf85,ok94}
\be \label{epsilon}
\epsilon(r) = D-{D-2\over 2}\,\left[ (sr)^2+2sr+2\right]\,e^{-sr}\ .
\ee
Empirical values for the parameters $D$ and $s$ are chosen, so that
for large distances the permittivity takes the value of bulk water,
$\epsilon =80$, and the value $\epsilon=2$ for short distances, i.e.
for the interior of the molecule. Approximating solvation effects in 
this way is implemented as an option in SMMP. It allows to include 
solvation effects without any significant slowing down over the vacuum 
simulation with $\epsilon =2$. The approach is clearly an 
oversimplification, because atoms which are close to each other do not 
necessarily have to be simultaneously in the interior of the molecule.
Reversely, two atoms which are separated by a large distance may still 
be in the interior of the molecule. More elaborated approaches are
asked for.

\begin{table}[ht]
\caption{ Atomic solvation parameter sets implemented in SMMP. The
first column gives the value of the SMMP parameter {\tt itysol}
and the second column the letter code used in SMMP. In the author 
column we give also the year of publication. The last column
indicates the method used as explained in the text.
\label{tab_ASP} }
\medskip
\centering
\begin{tabular}{|c|c|c|c|}   \hline
  &      & Authors                                        &    \\ \hline
1 & OONS & Ooi, Obatake, Nemethy, Scheraga~\cite{oons} 1987 & v/w\\ \hline
2 & JRF  & Vila, Williams, V\'asquez, Scheraga~\cite{jrf} 1991 & v/w \\ \hline
3 & WE92 & Wesson and Eisenberg~\cite{we92} 1992 & v/w \\ \hline
4 & EM86 & Eisenberg and McLachlan~\cite{em86} 1986 & o/w \\ \hline
5 & SCH1 & Eisenberg, Wesson, Yamashita~\cite{sch1} 1989 & o/w \\ \hline
6 & SCH2 & Kim~\cite{sch2}, see~\cite{jeh95} 1990 & o/w \\ \hline
7 & SCH3 & Wesson and Eisenberg~\cite{we92} 1992 & v/w \\ \hline
8 & SCH4 & Schiffer, Caldwell, Kollman, Stroud~\cite{sch4} 1993&v/ws\\ \hline
9 & BM   & Freyberg, Richmond, Brown~\cite{bm} 1993 & cla \\ \hline
\end{tabular} \end{table} 

   If the molecule-solvent interaction is proportional to the surface 
area of the atomic groups, it is given by the sum of contributions of 
a product of the surface area of each atomic group and the atomic
solvation parameter~\cite{em86},
\be \label{Esol}
    E_{\rm sol}=\sum_i \sigma_i A_i \, .
\ee
Here $E_{\rm sol}$ is the solvation energy and the sum is over all
atomic groups.  $A_i$ is the solvent accessible surface area and 
$\sigma_i$ the atomic solvation parameter of group $i$. The choice 
of a set of ASPs $\sigma_i$ defines a model of solvation. There are 
nine sets of ASPs in the SMMP package, which we list in 
Table~\ref{tab_ASP}. Columns one and two of this table gives the 
notations used in SMMP to identify the different sets.

Eisenberg and McLachlan~\cite{em86} were the first to determine a set
of ASPs ({\tt itysol}$=4$, EM86 in the SMMP notation). For this, they 
considered the process of transferring atoms or groups of atoms from 
the interior of a protein to aqueous solution and used transfer 
energies of amino acids from $n$-octanol to water as reported 
in~\cite{fp83}. The ASPs are then determined by 
least-square fitting. Octanol is chosen, because it apparently 
resembles the interior of a protein. With the exception 
of~\cite{sch4} and~\cite{bm}, the other authors used similar methods 
with the major variation that instead of transfer energies with 
respect to octanol-water (o/w) also transfer energies with respect 
to vacuum-water (v/w) were used (for early determination of v/w 
transfer energies see~\cite{c81,w81}). The last column of 
Table~\ref{tab_ASP} indicates whether the transfer energy is 
o/w or v/w. In the chronological order~\cite{em86,sch1,we92} Eisenberg 
and collaborators contributed the parameter sets EM86, SCH1, WE92
and SCH3. Scheraga and collaborators~\cite{oons,jrf} contributed the 
parameter sets OONS and JRF. Here it should be noted that some of the 
original ASP sets were modified in course of time. For 
{\tt itysol}$=1,\dots,8$ SMMP implements the parameters as reviewed 
and tabulated in Ref.~\cite{jeh95}, where in turn the sets SCH1 to 
SCH4 are simply taken from Schiffer et al.\cite{sch4}. Table~1 of 
SMMP~\cite{smmp} lists the implemented ASPs for ${\tt itysol}=1,\dots,8$.

Somewhat special cases are the ASP sets SCH4~\cite{sch4} and BM~\cite{bm}.
SCH4 was determined by comparison of the crystal structure from molecular 
dynamics simulations of small peptides and proteins in explicit water with 
similar simulations using an ASP solvation term (v/ws). The BM set of 
SMMP relies on a specific classification (cla) of atomic groups, where for 
all non-hydrogen atoms the solvation coefficients are set to 1$\,$kcal/mol
per \AA$^2$. 

\subsection{Metropolis methods} \label{sec_met}

For the updating of our systems we use PT with two processors, one 
running at 300$\,$K and the other at 400$\,$K. This builds on the 
experience~\cite{b03} with vacuum simulations of Met-Enkephalin for 
which the following observations are made:

\begin{enumerate}
\item The integrated autocorrelation time $\tau_{\rm int}$ (defined
below in this section) increases from 400$\,$K to 300$\,$K by a factor
of ten for the (internal) energy and by factors of more than twenty for
certain dihedral angles.
\item The energy probability densities (pds) at 300$\,$K and 400$\,$K 
overlap sufficiently, so that the PT method works, leading to an 
improvement factor of about 2.5 in the real time needed for the 
simulation (see Table~I of Ref.~\cite{b03}).
\end{enumerate}

A brief description of the PT algorithm is given in the 
following. PT performs $n$ canonical MC simulations at different 
$\beta$-values with Boltzmann weight factors
\be
  w_{B,i}(E^{(k)})=e^{-\beta_i E_i^{(k)}}=e^{-H}\, ,\, i=0,\ldots,n-1
\ee
where $\beta_0<\beta_1<\ldots<\beta_{n-2}<\beta_{n-1}$
and a configuration is denoted by $k$. PT allows the exchange of 
neighboring $\beta$-values
\be
 \beta_{i-1} \leftrightarrow \beta_{i} \,\, {\rm for} \,\, i=1,\ldots,n-1 \; .
\ee
These transitions lead to the change
\begin{eqnarray} \nonumber
   -\Delta H &=& -\beta_{i-1}\,\left(E_i^{(k)}-E_{i-1}^{(k')}\right)
                 -\beta_i\,\left(E_{i-1}^{(k')}-E_i^{(k)}\right)\\
 &=& (\beta_i-\beta_{i-1})\,\left(E_i^{(k)}-E_{i-1}^{(k')}\right)
\end{eqnarray}
which is accepted or rejected according to the Metropolis algorithm,
i.e. with probability one for $\Delta H \leq 0$ and with
probability $\exp(-\Delta H)$ for $\Delta H > 0$.

For the vacuum system the performance of the PT simulation was 
improved by an additional factor of two in Ref.\cite{b03} by using 
a first approximation, called RM$_1$, to the rugged Metropolis scheme 
introduced there. A (short) simulation at 400$\,$K was used to obtain 
estimates $\overline\rho_j(v_j)$, $j=1,\dots,24$ of the pds of the 24 
dihedral angles, which were then fed into the simulation. For a 
configuration change $k\to k'$ at temperature $T_i$ the new 
configuration accepted with the probability
\be \label{RM1}
p_{\rm acpt} = \min\left[1,{
\exp\left(-\beta_i\,E_i^{(k')}\right)\,
\prod_{j=1}^{24} \overline\rho_j \left(v_j^{(k')}\right) \over
\exp\left(-\beta_i\,E_i^{(k)}\right)\, 
\prod_{j=1}^{24} \overline\rho_j \left(v_j^{(k)}\right) }
\right]
\ee
in the RM$_1$ updating scheme. In the present paper we report the 
improvement due to this biased updating for some of the ASP sets.

\begin{figure}
  \begin{center}
\psfig{file=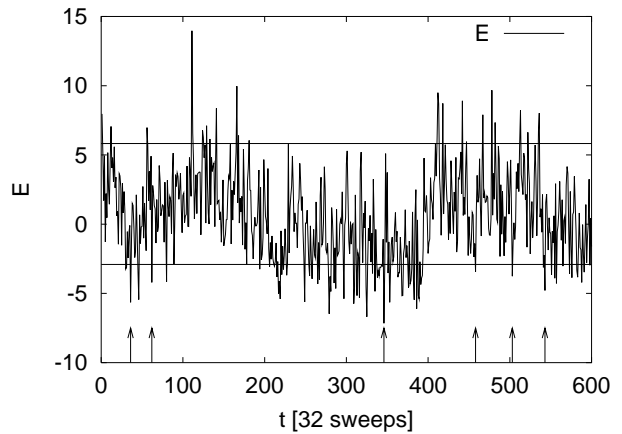,width=8.4cm, angle=0}
   \caption{Selection of configurations for local minimization
    from the energy time series at 300$\,$K. The lower and upper 
    straight lines indicate the quantiles $E_{0.1}$ and $E_{0.9}$, 
    respectively.}
    \label{fig_time_E}
  \end{center}
\end{figure}

In the vacuum simulation it is possible to determine the GEM by 
minimizing selected configurations of the 300$\,$K time series. 
Here we apply the same procedure to our PT simulation of the ASP
sets introduced in the previous subsection: 

\begin{enumerate}
\item We determine the lower 10\% quantile $E_{0.1}$ and the upper
10\% quantile $E_{0.9}$ of the energy distribution of our time 
series. This is done by sorting all energies in increasing order
and finding the values which cut out the lower and upper 10\% of
the data. For the statistical concepts see, e.g., Ref.~\cite{brandt}.

\item We partition the time series into bunches of configurations.
A bunch contains the configurations from one crossing of the
upper quantile $E_{0.9}$ to the next so that at least on crossing
of the lower quantile $E_{0.1}$ is located between the two crossings
of $E_{0.9}$. For each bunch we pick then its configuration of 
lowest energy. The idea behind this procedure is to pick minima of 
the time series, which are to a large degree statistically independent.
In Fig.~\ref{fig_time_E} the arrows indicate the energy values
picked in that way from the first 600 configurations recorded in
the RM$_1$ simulation of Ref.~\cite{b03}.

\item We run a conjugate gradient minimizer on all the selected
configurations and thus obtain a set of configurations which are 
local energy minima. For the vacuum simulation~\cite{b03} about
5\% to 6\% of the thus minimized configurations agree with the
GEM.
\end{enumerate}

To determine the speed at which the systems equilibrate, we measure the
integrated autocorrelation time $\tau_{\rm int}$ for the energy and each 
dihedral angle. In particular the integrated autocorrelation times are 
directly proportional to the computer run times needed to achieve the
same statistical accuracy for each system. They thus determine the 
relative performance of distinct algorithms. For an observable $f$ the 
autocorrelations are 
\be
 C(t) = \langle f_0\,f_t\rangle - \langle f\rangle^2
\ee
where $t$ labels the computer time. Defining $c(t)=C(t)/C(0)$, the
time-dependent integrated autocorrelation time is given by
\be \label{tau_int_t}
 \tau_{\rm int}(t) = 1 + 2 \sum_{t'=1}^t c(t')\ .
\ee
Formally the integrated autocorrelation time $\tau_{\rm int}$ is 
defined by $\tau_{\rm int}=\lim_{t\to\infty}\tau_{\rm int}(t)$.
Numerically, however, this limit cannot be reached as the noise 
of the estimator increases faster than the signal. Nevertheless,
one can calculate reliable estimates by reaching a window of $t$ 
values for which $\tau_{\rm int}(t)$ becomes flat, while its error 
bars are still reasonably small. This is the method we employ in the
next section, see Ref.~\cite{sokal} for a more detailed discussion
of the integrated autocorrelation time.

\section{Results} \label{results}

\subsection{Autocorrelations}

The PT simulations with temperatures $T_0=400\,$K and $T_1=300\,$K are 
performed on the system in vacuum ($\epsilon=2$), with $\epsilon(r)$ 
given by Eq.~(\ref{epsilon}) and for the nine ASP sets of 
Table~\ref{tab_ASP}. The dihedral angles updated in our simulations 
are fully variable in the range from $-\pi$ to $\pi$. We keep a time 
series of $2^{16}=65536$ configurations for each replica (i.e., each 
of the two processor), in which subsequent configurations are 
separated by 32 sweeps. A sweep is defined by updating each dihedral 
angle once sequentially. Before starting with measurements
$2^{18}=262144$ sweeps are performed for reaching equilibrium.
Thus, the entire simulation at one temperature relies on
$2^{21}+2^{18}=2,359,296$ sweeps.
On the Cray T3E, this takes about 14 hours for the vacuum system
and $5\times 14$ hours for each ASP set.

\begin{table*}
\caption {\label{tab_r1} Average energies $\langle E\rangle\,$(Kcal/mol), 
acceptance rates and integrated autocorrelations times $\tau_{\rm int}$ 
for the energy are shown for simulations in vacuum (VAC), with 
$\epsilon(r)$ of Eq.~(\ref{epsilon}) and with the nine ASP introduced
in Table~\ref{tab_ASP}.}
\begin{ruledtabular}
\begin{tabular}{c|rrr|rrr}
\multicolumn{1}{c|}{Set}       &  \multicolumn{3}{c|}{$T=400$ K} & 
\multicolumn{3}{c}{$T=300$ K}      \\
\hline
       &  \multicolumn{1}{c}{$\langle E\rangle$} &\multicolumn{1}{c}{acpt}&
\multicolumn{1}{c|}{$\tau_{\rm int}$} &\multicolumn{1}{c}{$\langle E\rangle$}&
\multicolumn{1}{c}{acpt}&\multicolumn{1}{c}{$\tau_{\rm int}$} \\
\hline
 VAC   &    7.07(03) & 0.167 & 3.67(20) &    1.29(06) & 0.119 & 19.9(1.6) \\
$\epsilon(r)$ 
       &  -12.00(03) & 0.171 & 2.92(10) &  -17.61(06)& 0.121 & 14.35(75)\\
 OONS  &  -13.80(01) & 0.195 & 1.25(02) &  -17.70(02) & 0.143 & 2.64(14) \\
 JRF   & -311.69(44) & 0.058 & -        & -319.08(40) & 0.046 & - \\
 WE92  &  -15.76(02) & 0.199 & 1.30(03) &  -19.75(02) & 0.145 & 2.94(07) \\
 EM86  &   13.49(03) & 0.158 & 4.71(21) &    8.03(06) & 0.116 & 25.0(2.9) \\
 SCH1  &   10.45(03) & 0.165 & 3.72(22) &    4.95(06) & 0.119 & 23.2(2.2) \\
 SCH2  &  -18.33(02) & 0.212 & 1.11(01) &  -21.83(01) & 0.160 & 1.89(05) \\
 SCH3  &   13.33(03) & 0.151 & 4.59(34) &    8.33(06) & 0.112 & 26.4(3.3) \\
 SCH4  &   13.38(03) & 0.158 & 4.35(17) &    7.85(05) & 0.115 & 25.1(2.1) \\
 BM    &  630.4(3.9) & 0.043 & -        &  610.6(3.0) & 0.037 & - \\
\end{tabular}
\end{ruledtabular}
\end{table*}

\begin{figure}
  \begin{center}
\psfig{file=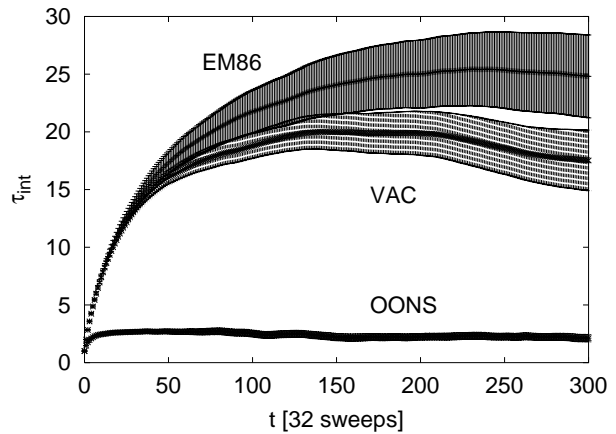,width=8.4cm, angle=0}
   \caption{ The time-dependent integrated autocorrelation time
   for the energy at 300$\,$K from our simulations of the vacuum 
   system and two of the solvent models of Table~\ref{tab_ASP}. } 
   \label{fig_taug}
  \end{center}
\end{figure}

Results of the average energy, acceptance rates and integrated 
autocorrelations times for the energy $E$ variable are shown in 
Table~\ref{tab_r1}. For the vacuum simulations and the ASP sets 
OONS and EM86 the time-dependent integrated autocorrelations 
times~(\ref{tau_int_t}) are shown in Fig.~\ref{fig_taug}. In 
each case a window of $t$ values is reached for which 
$\tau_{\rm int}(t)$ does no longer increase within its statistical
errors. In the case of the vacuum simulations it even does decrease,
but this is not significant due to the statistical error. These 
windows are then used to estimate the asymptotic $\tau_{\rm int}$
values of Table~\ref{tab_r1}. With the exception of the ASP sets
JRF and BM, the integrated autocorrelations times of all other 
sets are determined in the same way.

\begin{figure}
  \begin{center}
\psfig{file=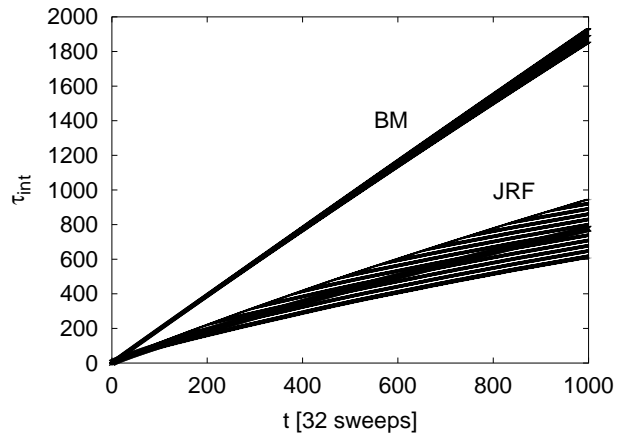,width=8.4cm, angle=0}
   \caption{The time-dependent integrated autocorrelation time for
    the energy at 300$\,$K from our simulations of the solvent 
    models JRF and BM.} \label{fig_taub}
  \end{center}
\end{figure}

From Table~\ref{tab_r1} we see that the acceptance rates of the solvent 
models JRF and BM are much lower than for the other models. In essence 
the simulations of these two models get stuck, which implies that their 
integrated autocorrelation times cannot be measured. This is illustrated 
in Fig.~\ref{fig_taub} for the time-dependent integrated autocorrelation 
time of the energy at 300$\,$K. The function $\tau_{\rm int}(t)$ increases 
rapidly until it gets lost in the noise. The pds of the dihedral angles
of these two models are also erratic and the conclusion is that they
cannot be used to describe Met-Enkephalin in solvent.

\begin{figure}
  \begin{center}
\psfig{file=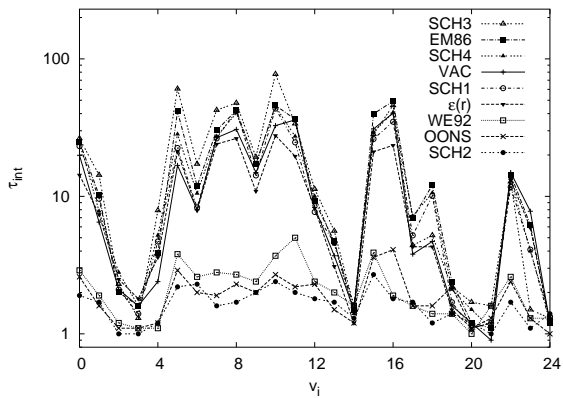,width=5.4cm, angle=270}
   \caption{Integrated autocorrelation times for the energies 
   ($v_0=E$) and the dihedral angles $v_i$, $i=1,\dots,24$ at 
    T=300$\,$K. The up-to-down order of the curves agrees at $i=10$
    with the order in the figure legend.} \label{fig_tau_int}
  \end{center}
\end{figure}

\begin{table}[ht]
\caption{ Definitions of the dihedral angles together with their 
integrated autocorrelation times $\tau_{\rm int}$ at 300$\,$K for 
simulations of WE92 with a statistics reduced by 1/8 and configurations 
recorded every four sweeps. PT denotes the $400\,$K$-300\,$K 
parallel tempering simulation. For PT-RM$_1$ the PT simulation
is supplemented by the RM$_1$ biased updating (\ref{RM1}) with 
input pds from 400$\,$K.
The factor of the last column denotes the increase of the PT
$\tau_{\rm int}$ over its values for the full WE92 simulations 
where configurations were recorded every 32 sweeps (8 is the 
upper bound for this factor).  \label{tab_dia} }
\medskip
\centering
\begin{tabular}{||c|c|c|c|c|c|c||}                   \hline
var     & angle  &res~\cite{ls87,mmm94}&res~\cite{smmp}
                            &PT-RM$_1$& PT    & factor \\ \hline
$v_1$   &$\chi^1$&Tyr-1&Tyr-1&  6.9 (1.1)& 11.6 (1.6)&  6.1 (0.9)\\ \hline
$v_2$   &$\chi^2$&Tyr-1&Tyr-1&  2.0 (0.2)&  3.1 (0.5)&  2.7 (0.5)\\ \hline
$v_3$   &$\chi^6$&Tyr-1&Tyr-1&  1.0 (0.1)&  1.3 (0.2)&  1.3 (0.2)\\ \hline
$v_4$   &$\phi$  &Tyr-1&Tyr-1&  2.1 (0.2)&  2.6 (0.4)&  2.4 (0.4)\\ \hline
$v_5$   &$\psi$  &Tyr-1&Gly-2& 12.6 (1.7)& 15.7 (2.2)&  4.1 (0.7)\\ \hline
$v_6$   &$\omega$&Tyr-1&Gly-2&  3.9 (0.4)& 14.4 (1.2)&  5.6 (0.5)\\ \hline
$v_7$   &$\phi$  &Gly-2&Gly-2&  9.1 (1.0)& 13.0 (1.4)&  4.6 (0.6)\\ \hline
$v_8$   &$\psi$  &Gly-2&Gly-3& 10.6 (1.3)& 20.4 (3.1)&  7.6 (1.2)\\ \hline
$v_9$   &$\omega$&Gly-2&Gly-3&  3.4 (0.2)& 16.0 (1.8)&  6.7 (0.8)\\ \hline
$v_{10}$&$\phi$  &Gly-3&Gly-3& 18.2 (3.2)& 31.0 (5.1)&  8.4 (1.5)\\ \hline
$v_{11}$&$\psi$  &Gly-3&Phe-4& 15.6 (2.9)&  52 (13)  &  12 (4) \\ \hline
$v_{12}$&$\omega$&Gly-3&Phe-4&  4.4 (0.6)& 17.7 (2.7)&  7.7 (1.3)\\ \hline
$v_{13}$&$\chi^1$&Phe-4&Phe-4&  3.3 (0.4)&  6.9 (1.1)&  4.4 (0.8)\\ \hline
$v_{14}$&$\chi^2$&Phe-4&Phe-4&  1.7 (0.2)&  3.2 (0.4)&  3.0 (0.4)\\ \hline
$v_{15}$&$\phi$  &Phe-4&Phe-4&  8.9 (1.3)& 19.6 (3.2)&  6.3 (1.2)\\ \hline
$v_{16}$&$\psi$  &Phe-4&Met-5&  4.5 (0.3)&  8.0 (0.9)&  4.4 (0.6)\\ \hline
$v_{17}$&$\omega$&Phe-4&Met-5&  1.8 (0.2)&  8.1 (1.2)&  5.4 (0.8)\\ \hline
$v_{18}$&$\chi^1$&Met-5&Met-5&  2.7 (0.2)&  8.3 (2.5)&  6.3 (1.9)\\ \hline
$v_{19}$&$\chi^2$&Met-5&Met-5&  1.9 (0.2)&  5.3 (0.4)&  4.0 (0.5)\\ \hline
$v_{20}$&$\chi^3$&Met-5&Met-5&  1.1 (0.1)&  2.7 (0.2)&  2.5 (0.2)\\ \hline
$v_{21}$&$\chi^4$&Met-5&Met-5&  1.0 (0.1)&  1.3 (0.1)&  1.3 (0.1)\\ \hline
$v_{22}$&$\phi$  &Met-5&Met-5& 36 (18)   & 23.8 (5.6)&  9.5 (2.4)\\ \hline
$v_{23}$&$\phi$  &Met-5&Met-5&  1.4 (0.2)&  1.9 (0.1)&  1.9 (0.1)\\ \hline
$v_{24}$&$\omega$&Met-5&Met-5&  1.0 (0.1)&  3.4 (0.2)&  3.1 (0.4)\\ \hline
$v_0$   &$E$     &     &     &  9.0 (1.7)& 19.4 (3.1)&  6.6 (1.1)\\ \hline
\end{tabular} \end{table} \vspace*{0.2cm}

The energy couples to all dihedral angles and its integrated 
autocorrelation time is characteristic for the entire system, while
the integrated autocorrelation times of the single dihedral angles 
vary heavily from angle to angle. For all our systems, but JRF and BM,
we show in Fig.~\ref{fig_tau_int} the integrated autocorrelation times 
at 300$\,$K for the energy and all dihedral angles. The notation 
$v_i$, $i=0,1,\dots,24$ is used, where $v_0$ is stands in for the
energy $E$ and for $i=1,\dots ,24$ the $v_i$ are the dihedral angles 
as used in the SMMP computer program. The relationship of the $v_i$
angles to the conventional notation for dihedral angles and their 
residues is summarized in Table~\ref{tab_dia}, where it should be 
noted that the SMMP notation~\cite{smmp} differs from other
literature~\cite{ls87,mmm94}.

In Fig.~\ref{fig_tau_int} we see that for each dihedral angle $v_i$ 
the integrated autocorrelation times $\tau_{\rm int}[v_i]$ for the 
three solvent models OONS, WE92 and SCH2 are smaller than for the 
remaining systems, including the vacuum system. For the integrated 
autocorrelation time of the energy $\tau_{\rm int}[E]$ this 
observation is already obvious from Table~\ref{tab_r1}. In
particular, this means that the OONS, WE92 and SCH2 models require 
far less statistics than the vacuum run for achieving the same 
accuracy of results. Using the $\tau_{\rm int}[E]$ results of 
Table~\ref{tab_r1}, we find a factor in the range 7 to 10, which 
more than offsets the factor of 5 by which the ASP model 
simulations are slower than the vacuum simulation. In the following
the solvation models OONS, WE92 and SCH2 define the ``fast class'',
while the other models shown in Fig.~\ref{fig_tau_int} constitute
the ``slow class'' (the models JRF and BM are omitted from this
classification). ``Good" behavior of the models OONS and WE92
has been previously observed~\cite{ok98}.

The autocorrelation times in the fast class are so 
small that the resolution of 32 sweeps in our recorded time series 
becomes too crude. Namely, autocorrelations over less than
32 sweeps are then not measured and the integrated autocorrelation
time approaches one as soon as autocorrelations stay within the 
range of 32 sweeps. To investigate this point further, we performed 
for the OONS, WE92 and SCH2 models simulations for which the 
configurations were recorded every four sweeps and the total 
statistics was reduced by the factor 1/8. In the new units of four
sweeps the integrated autocorrelation time is larger by a factor 
which is bounded by $8=32/4$. The bound is assumed, if there is no
improvement due to integrating additional small fluctuations out 
(i.e. due to the additional configuration in between the 32 sweeps,
which are now kept in the time series). 

For WE92 we report in the PT column of Table~\ref{tab_dia} the 
integrated autocorrelation times from the simulation with reduced 
statistics. For many dihedral angles the increase lies well below 
the factor of eight, showing that we gain in accuracy by averaging 
over small fluctuation within the range of 32 sweeps. On the 
other hand, nothing is gained by this extra averaging for several 
angles with large autocorrelations. In those case the simulations
yield within their statistical errors the upper bound 8. 

To supplement the vacuum results of Ref.~\cite{b03}, we repeated 
the WE92 PT simulations by using estimates of the dihedral pds 
from 400$\,$K as input for the biased updating of Eq.~(\ref{RM1}). 
These results are reported in the PT-RM$_1$ column of 
Table~\ref{tab_dia}. As in the case of the vacuum simulations, we 
find an improvement of the PT performance by a factor of 
approximately two, which is also obtained for the other models 
of the fast class.  For the slow class we checked by direct 
improvement of the original simulations on the ASP models 
EM86 and SCH4 and find again an acceleration by a factor 
of about two when we are using RM$_1$ updating.

\subsection{Structure}

\begin{table*}
\caption {\label{tab_r2} Determination of local minima: $E_{0.1}$ and 
$E_{0.9}$ are the lower and upper 10\% quantiles of the energies of the 
time series recorded, $N_{\rm conf}$ denotes the number of configurations 
prepared for further minimization, $E_{\rm min}\,$(Kcal/mol) is the lowest 
energy found, and $N_{\rm hits}$ is the number of times the lowest energy 
configuration was hit. }
\begin{ruledtabular}
\begin{tabular}{c|rrrrr|rrrrr}
\multicolumn{1}{c|}{Set} &\multicolumn{5}{c|}{$T=400$ K} &
\multicolumn{5}{c}{$T=300$ K}    \\
\hline
  &\multicolumn{1}{c}{$N_{\rm conf}$} &\multicolumn{1}{c}{ $E_{0.1}$} &
\multicolumn{1}{c}{$E_{0.9}$} &\multicolumn{1}{c}{$E_{\rm min}$} &
\multicolumn{1}{c|}{$N_{\rm hits}$}&\multicolumn{1}{c}{$N_{\rm conf}$}
 &\multicolumn{1}{c}{$E_{0.1}$} &\multicolumn{1}{c}{$E_{0.9}$} & 
\multicolumn{1}{c}{$E_{\rm min}$} &\multicolumn{1}{c}{$N_{\rm hits}$} \\
\hline
VAC  & 2190 & 1.98 & 12.26 & -12.91 & 13 & 1073  & -2.98 & 5.73 &  -12.91 & 55 \\
$\epsilon(r)$  &  2622  & -16.95 & -6.97 & -31.94& 8 &  1312  & -21.85 & -13.17 & -31.94   & 27\\
OONS & 3315 & -17.83 & -9.63 & -27.69 &  1 & 2641 & -21.40 & -13.96 & -28.93  &  1 \\ 
JRF  &  448 &  -317.96 & -304.98 &-328.72& 1 & 365  & -323.66 & -314.24 & -332.87  &  1 \\
WE92 & 3307 &  -19.88 & -11.52 & -29.44 &  1 &  2453  & -23.43 & -15.95 & -30.39  &  1 \\
EM86 & 2307 &  8.57 & 18.59 & -4.11&   1 & 1191  & 3.83 & 12.39 & -5.47  &  1 \\
SCH1 & 2511 &  5.57 & 15.45 & -5.54&   1 & 1147  & 0.71 & 9.33 &   -7.52  &  1 \\
SCH2 & 3454 & -22.17 & -14.34 &-31.32& 1 &  2918 & -25.31 & -18.32 &  -32.71   &  1 \\
SCH3 & 2315 & 8.57 & 18.32 &-1.70&  1&  1229  & 4.29 & 12.50 & -3.29    &  1 \\
SCH4 & 2331 & 8.45 & 18.50 & -4.93&  1&  1108  & 3.66 & 12.23 &  -5.16    &  1 \\  
BM   & 2  & 606.37 & 655.16 & 594.78& 1  & 1 & 598.37 & 646.35 & 590.32    &  1 \\
\end{tabular}
\end{ruledtabular}
\end{table*}

For all our simulations we applied the method outlined in 
subsection~\ref{sec_met} to determine local energy minima and some
results are summarized in Table~\ref{tab_r2}: $E_{0.1}$ and 
$E_{0.9}$ are the lower and upper 10\% quantiles of the energy
and $N_{\rm conf}$ denotes the number of minima of the time series
prepared for further minimization. The lowest energy found in 
this minimization process is denoted by $E_{\min}$ and $N_{\rm hits}$
is the number of times the lowest energy configuration was hit. 
While the absolute values of $E_{0.1}$ and $E_{0.9}$ vary 
considerably from set to set, the differences $E_{0.9}-E_{0.1}$ 
stay similar. The explanation is that the ASP sets differ by large 
additive constants to the energy.

Again, the results of the JRF and BM solvent models are erratic. The
BM model is entirely frozen, $N_{\rm conf}=2$ at 400$\,$K and 
$N_{\rm conf}=1$ at 300$\,$K. Therefore, we do not give minimization
results for BM. For JRF the $N_{\rm conf}$ numbers are more reasonable,
but still by a factor of one third and less smaller than the
$N_{\rm conf}$ numbers of each other system. JRF is also disregarded
in the following discussion.

Only if $N_{\rm hits}>1$ holds we have an indication that we found
the GEM. Interestingly, this happens for none of the ASP solvent 
models, while it is the case for the vacuum and the $\epsilon(r)$
simulations (notably already at 400$\,$K). 
Quite some time ago Li and Scheraga~\cite{ls87,ls88} 
developed a Monte Carlo minimization method and applied it to 
Met-Enkephalin in vacuum and in solvent modeled by OONS. While 
for the vacuum system their method converged consistently to the 
GEM, all their five runs of the solvent model led to different 
conformations with comparable energies. They interpreted their 
results in the sense that Met-Enkephalin in water at $20^{\circ}$C 
is likely in an unfolded state for which a large ensemble of 
distinct conformations co-exist in equilibrium. A consistent
scenario was later observed in NMR experiments~\cite{grcahi}.

Although the 
minimization method of  Li and Scheraga is entirely different 
from ours, they essentially tested for valleys of attraction 
to the GEM at room temperature, quite as we do in the present 
paper. So, we have not only confirmed their old result, but find 
that it is also common to a large set of ASP models implemented in 
SMMP. Neither the method by which an ASP set was derived, nor 
whether it belongs to the fast or slow class, appears to matter 
with this respect. 

\begin{figure}
  \begin{center}
\psfig{file=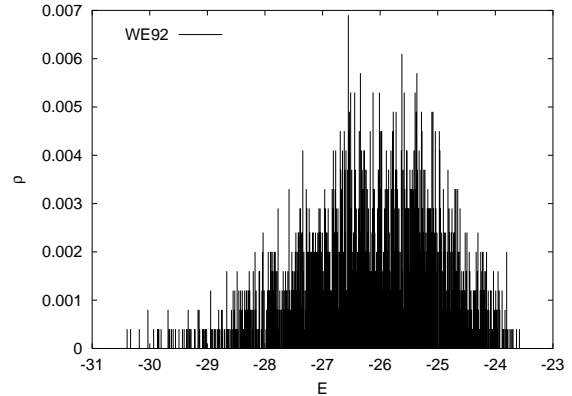,width=5.4cm, angle=270}
   \caption{Local energy minima for the WE92 solvation model
   as obtained by our minimization method.}
  \label{fig_Emin_we92}
  \end{center}
\end{figure}

To give one example, the frequency of local energy minima of the WE92 
solvation model as obtained by our minimization procedure from the 
300$\,$K time series is depicted in Fig.~\ref{fig_Emin_we92}. 
$N_{\rm conf}=2453$ minimizations are performed. Our lowest energy 
state is only found once and the same holds for the close-by low energy 
states. Fig.~\ref{fig_Emin_we92} should be compared with Fig.~2 of 
Ref.~\cite{b03}, where the frequency of the low energy minima of the 
vacuum simulation is shown. There the lowest energy state relies on 
107 entries out of 1913 minimizations~\cite{errata}.

\begin{figure}
  \begin{center}
\psfig{file=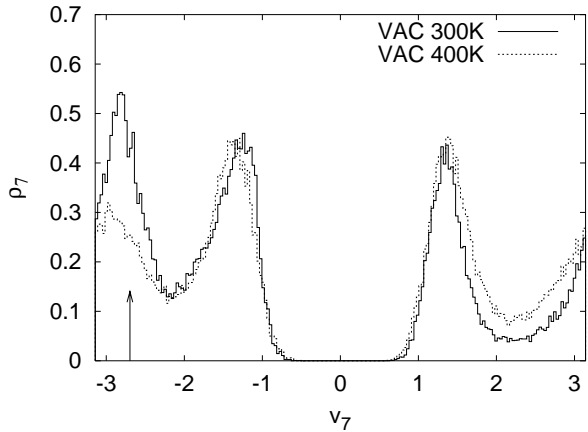,width=8.4cm, angle=0}
   \caption{Probability density of the dihedral angle $v_7$ for
    the vacuum simulation. The arrow indicates the vacuum GEM
    value of this angle. } \label{fig_v7_vac}
  \end{center}
\end{figure}

\begin{figure}
  \begin{center}
\psfig{file=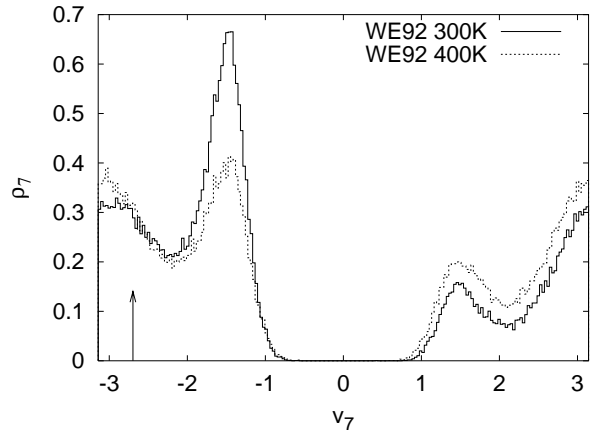,width=8.4cm, angle=0}
   \caption{Probability density of the dihedral angle $v_7$ for
    the WE92 simulation. The arrow indicates the vacuum GEM 
    value of this angle.} \label{fig_v7_we92}
  \end{center}
\end{figure}

In a search for structural differences of Met-Enkephalin in
vacuum, or in the $\epsilon(r)$ system, versus the ASP models, 
we looked at the pds of the dihedral angles
at 300$\,$K. For all systems together there are $9\times 24 = 216$ 
figures to consider. At the first look the pds of the different
systems are amazingly similar, independently of whether they are 
from systems of the fast or slow class, from an ASP model, from the 
vacuum or from the $\epsilon(r)$ simulation.
A more careful investigation reveals differences, which appear 
to relate to the distinct behavior under our minimization. For the 
dihedral angle $v_7$ this is illustrated in Fig.~\ref{fig_v7_vac} and
Fig.~\ref{fig_v7_we92}. Its probability densities are compared at 
300$\,$K and  400$\,$K. For the vacuum simulation the pds are 
depicted in Fig.~\ref{fig_v7_vac} and from 400$\,$K to 300$\,$K 
we observe an increase of the peak which is located close to
the arrow, which indicates the vacuum GEM value of $v_7$. In contrast
to this, the wrong peak increases in Fig.~\ref{fig_v7_we92}, where
the pds are shown for the WE92 solvent model.

\begin{figure}
  \begin{center}
\psfig{file=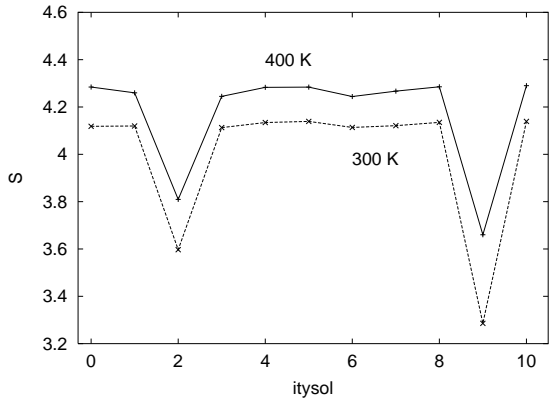,width=5.4cm, angle=270}
   \caption{Entropies of the pds of our ASP models. The models are 
   labeled according Table~\ref{tab_ASP}, in addition 
   ${\tt itysol}=0$ for vacuum and ${\tt itysol}=10$
   for the $\epsilon(r)$ model.} \label{fig_entropy}
  \end{center}
\end{figure}

One may suspect that the difference between the models of our fast
and slow class is simply due to an effectively higher temperature
for the three models of the fast class. To gain insight into this
question, we calculate the entropies of our pds. Each pd is 
discretized as a histogram of 200 entries, $\rho_{ij}$, where
$i=1,\dots,24$ labels the dihedral angles according to 
Table~\ref{tab_dia} and $\sum_{j=1}^{200}\rho_{ij}=1$. The 
entropy of the pd of a dihedral is then defined by
\be \label{entropy}
S_i = - \sum_{j=1}^{200} \rho_{ij}\,\ln \rho_{ij}
\ee
and the total entropy of the pds of an APS model is $S=\sum_iS_i$.
In Fig.~\ref{fig_entropy} the thus obtained entropies are depicted
for all our models. The lines between the data points have no other
meaning but to guide the eyes. The dips for the JRF and the BM model 
show, again, that their configurations are essentially frozen. For 
the other we see a decrease of entropy from 400$\,$K to 300$\,$K,
but we find no larger entropy for the models of the fast class than
for the models of the slow class. Therefore, the effective 
temperature scenario is rules out. Instead, it seems that for
the models of the fast class the solvent has some kind of 
``lubrication'' effect, which accelerates the simulation.

\begin{table*}
\caption {\label{tab_Esol} Transfer energies.}
\begin{ruledtabular}
\begin{tabular}{|c|rr|rr|rr|rr|}
\multicolumn{1}{|c|}{Set} &  \multicolumn{6}{c|}{$T=300$ K} 
                         & \multicolumn{2}{c|}{$T=400$ K} \\ \hline
& \multicolumn{2}{c|}{$\langle E^T\rangle$(Local min.)} 
& \multicolumn{2}{c|} {$\langle E^T\rangle$(Time series min.)}  
& \multicolumn{2}{c|}{$\langle E^T\rangle$} 
& \multicolumn{2}{c|}{$\langle E^T\rangle$} \\ \hline
     &  s/v    & $-$v/s  &  s/v    & $-$v/s  &   s/v   & $-$v/s  &   s/v   &
 $-$s/v \\  \hline
OONS & $-$24.85& $-$18.71& $-$24.63& $-$18.53& $-$24.84& $-$19.07& $-$24.94&
  -20.19 \\
JRF  &$-$346.95&$-$185.65&$-$346.72&$-$187.30&$-$343.56&$-$196.96&$-$337.68&
 -209.46 \\
WE92 & $-$28.84& $-$17.46& $-$28.52& $-$17.62& $-$27.87& $-$18.76& $-$27.47&
 $-$20.61 \\
EM86 &     6.19&     6.93&     6.29&     7.01&     6.45&     7.08&     6.60&
     7.11 \\ 
SCH1 &     2.76&     3.89&     2.86&     3.92&     2.94&     3.80&     2.97&
     3.60 \\
SCH2 & $-$31.03& $-$20.61& $-$30.77& $-$20.80& $-$30.61& $-$21.97& $-$30.45&
 $-$23.85 \\
SCH3 &     4.03&     9.73&     4.28&     9.80&     4.74&     9.50&     5.15&
     9.00 \\
SCH4 &     6.08&     6.86&     6.17&     6.93&     6.32&     6.95&     6.46&
     6.94 \\
BM   &  $-$    &   715.29&  $=$    &   721.43&   581.60&   745.22&   597.61&
   776.03 \\
\end{tabular} \end{ruledtabular} \end{table*}

\begin{table*}
\caption {\label{tab_gy_ee} Gyration radii and end to end distance.}
\begin{ruledtabular}
\begin{tabular}{|c|rr|rr|rr|rr|}
\multicolumn{1}{|c|}{Set} &  
\multicolumn{6}{c|}{$T=300$ K} & 
\multicolumn{2}{c|}{$T=400$ K} \\ \hline
       &  \multicolumn{2}{c|}{Local minima} &
\multicolumn{2}{c|}{Time series minima}  &
\multicolumn{2}{c|}{Time series} &
\multicolumn{2}{c|}{Time series} \\ \hline
    & $\langle R_{\rm gy}\rangle$
    & $\langle R_{\rm e-e}\rangle$
    & $\langle R_{\rm gy}\rangle$
    & $\langle R_{\rm e-e}\rangle$
    & $\langle R_{\rm gy}\rangle$
    & $\langle R_{\rm e-e}\rangle$
    & $\langle R_{\rm gy}\rangle$ 
    & $\langle R_{\rm e-e}\rangle$ \\
VAC & 4.56 & 5.67 & 4.60 & 5.83 & 4.72 & 6.83 & 4.97 & 8.50\\
$\epsilon{r}$ 
    & 4.53 & 6.20 & 4.57 & 6.33 & 4.71 & 7.39 & 4.99 & 8.94\\
OONS& 4.92 &10.30 & 4.94 &10.34 & 5.32 &11.71 & 5.60 &12.45\\
JRF & 5.75 &13.00 & 5.75 &13.00 & 5.78 &13.05 & 5.70 &13.20\\
WE92& 5.06 &12.06 & 5.07 &12.09 & 5.43 &13.02 & 5.72 &13.47\\
EM86& 4.47 & 6.97 & 4.48 & 7.02 & 4.62 & 7.94 & 4.86 & 9.28\\ 
SCH1& 4.51 & 6.96 & 4.52 & 7.02 & 4.68 & 7.96 & 4.96 & 9.45\\
SCH2& 5.18 &12.48 & 5.20 &12.53 & 5.63 &13.47 & 5.86 &13.66\\
SCH3& 4.54 & 8.88 & 4.55 & 8.90 & 4.72 & 9.56 & 4.95 &10.50\\ 
SCH4& 4.46 & 6.82 & 4.47 & 6.87 & 4.62 & 7.83 & 4.87 & 9.17\\   
BM  &  $-$ & $-$  & $-$  & $-$  & 4.13 & 7.30 & 4.21 & 7.66\\      
\end{tabular} \end{ruledtabular} \end{table*}

Strong similarities between the ASP models of the fast class on one side
and the ASP models of the slow class on the other side are found for the
solvation energies, the gyration radii and the end to end distances.

The solvation energies $E_{\rm sol}$ (\ref{Esol}) measured during our
simulations of ASP models are solvent-vacuum (s/v) transfer energies.
There are structural difference between the typical configurations of
an ASP model time series and the vacuum time series. Consequently, the 
average s/v transfer energies are not identical with the average
vacuum-solvent (v/s) transfer energies, which are obtained by 
calculating $E_{\rm sol}$ of the solvent models on the configurations 
of the vacuum time series. The s/v as well as the $-$v/s average transfer 
energies are collected in Table~\ref{tab_Esol}. The averages are taken
for the canonical time series at 300$\,$K and at 400$\,$K. At 300$\,$K
averages are also taken for the time series minima (indicated by
arrows in Fig.~\ref{fig_time_E}) and for the local minima (which are
obtained by running the conjugate gradient minimizer on the time
series minima). For the gyration radii $R_{\rm gy}$ and the end to
end distances $R_{\rm e-e}$ the same averages are collected in 
Table~\ref{tab_gy_ee} (definitions and software are given in SMMP).

For the transfer energies the over-all effect is hydrophilic for
the APS models of the fast class and hydrophobic for the APS models
of the slow class. Within each class the values are quite similar,
despite differences in the interaction coefficients (see Table~1
of Ref.\cite{smmp}). As expected the over-all transfer energies
of the JRF and BM models are out of the reasonable range, JRF to
the hydrophilic and BM to the hydrophobic side. Our Table~\ref{tab_gy_ee}
shows that we observe the extended structures found in previous 
simulations~\cite{ls88,koh98} and in NMR experiments~\cite{grcahi} only
for the APS models of the fast class.

\section{Summary and Conclusions} \label{conclusions}

We have performed Met-Enkephalin simulations at room temperature 
(300$\,$K) for the solvation models of Table~\ref{tab_ASP}. 
Quantitative results obtained in that way cannot be trusted, 
apparently because the methods to derive the ASPs have been quite 
crude. Also our simulations do not give information that would 
allow us to pick a best ASP set for the intended purpose of 
simulating Met-Enkephalin at 300$\,$K. Nevertheless, we obtain 
a qualitative overview of a number of interesting consequences,
which one can expect by including solvation effects via an ASP 
model in Metropolis calculations.

Two of the ASP sets (JRF~\cite{jrf} and BM~\cite{bm} as implemented
in SMMP~\cite{smmp}) suffer from so large 
autocorrelations that for them Metropolis simulations at 
300$\,$K are in essence impossible. Their dihedral angles are 
essentially frozen. These two ASP sets are certainly erratic, as 
300$\,$K is a temperature at which thermodynamic fluctuations of 
the systems are expected (also these two sets perform badly 
at 400$\,$K). 

The remaining nine models, seven ASP sets, $\epsilon=2$ vacuum, and
an $\epsilon(r)$ system~\cite{hrf85}, fall into a fast and a slow 
class with respect to their integrated autocorrelation times, 
see Fig.~\ref{fig_tau_int}.  Vacuum simulations are in the slow 
class. This leads to the 
interesting feature that it takes less computer time to estimate 
physical observables at room temperature in the fast solvation 
models OONS~\cite{oons}, WE92~\cite{we92}, and 
SCH2~\cite{sch2,sch4}, than it takes for 
vacuum, despite the substantial increase of the computer time 
per sweep by a factor of about 5 for the solvation models over 
the vacuum system. We have no clear clue why some models have a 
fast and others a slow dynamics. To derive the parameters of OONS 
and WE92 vacuum-water (v/w) transfer energies were used, but for 
SCH2 it was octanol-water (o/w). Also the slow class features v/w 
as well as o/w ASP models.

We applied the minimization procedure of Ref.~\cite{b03} in an
attempt to locate the GEM for the nine systems, which are 
reasonably well-behaved under Metropolis simulations at 300$\,$K.
The GEM is unambiguously found for the vacuum system and for
the simulation with a distance dependent electrostatic permittivity.
No true GEM is found for any of the remaining seven ASP models. This 
confirms an old result of Li and Scheraga~\cite{ls88}, who 
concluded that at room temperature Met-Enkephalin in water is 
likely in an unfolded state.
To get a better understanding of this result, we studied at
300$\,$K the dihedral pds in some details. At a first glance they
look quite similar for all the models in the fast as well as in the 
slow class. Differences are found for a number of details, which
may allow to explain why the 300$\,$K configurations of the ASP 
models behave entirely different under our minimization procedure 
than the vacuum and the $\epsilon(r)$ systems. 

The central question, which remains to be settled, is whether ASP 
models will ultimately allow for accurate Metropolis simulations 
of biomolecules like Met-Enkephalin in solvent or not. In principle,
this could be decided by determining whether ASPs exist which 
reproduce accurately mean energies of explicit solvent 
simulation around a large number of fixed Met-Enkephalin 
configurations.

\acknowledgments

We would like to thank Prof. P. Grassberger for useful discussions
and generous support of this work. The computer simulations were
carried out on the Cray T3E of the John von Neumann Institute for
Computing. BB acknowledges partial support by the U.S. Department 
of Energy under contract No. DE-FG02-97ER41022. In the final 
stage of this work we became aware of two articles where
ASP models are studied for the helix-coil transition of 
polyalanine~\cite{PeHa03}.

\end{document}